\documentclass[dvips,11pt]{article}
\usepackage{varioref,exscale,latexsym,amsmath,amssymb}
\usepackage{graphicx}
\usepackage{feynmp}
\def\beq{\begin{equation}}

\def\eeq{\end{equation}}

\def\beqa{\begin{eqnarray}}

\def\eeqa{\end{eqnarray}}

\title{{\bf The Nuclear Central Force in the Chiral Limit}}
\author{John F. Donoghue, \\ \\
Department of Physics\\
University of Massachusetts\\
Amherst, MA  01003, USA\\
and \\
Institut des Hautes \'{E}tudes Scientifiques \\
Bures sur Yvette, F-91440, France
 \\}
\begin{document}
\begin{titlepage}
\maketitle
\begin{abstract}
Chiral perturbation theory supplemented by the Omnes function is
employed to study the strength of the isoscalar central nuclear
interaction, $G_S$, in the chiral limit vs the physical case. A very
large modification is seen , i.e. $\eta_s = G_{S~{\rm chiral}}
/G_{S~{\rm physical}} = 1.37\pm 0.10 $.  This large effect is seen
to arise dominantly at low energy from the extra contributions made
by massless pions at energies near the physical threshold where the
physical spectral function must vanish kinematically. The slope away
from the chiral limit, $d_S$, is also calculated and is
correspondingly large. I also explain why this large variation is to
be expected.
\end{abstract}
\vspace{0.2 in}
\end{titlepage}

\section{Introduction}

For the most part we have obtained only a phenomenological
description of the nuclear force. Whether using meson exchange
potentials or effective field theory, the parameters of the
internucleon interaction are hard to relate directly to QCD. However
I will show that we know enough to understand the quark mass
dependence of the nuclear central interaction with reasonable
control. I will use this to describe the strength of the nuclear
interaction in the chiral limit. We will see that there exists a
strong variation of the strength, which nevertheless comes from
readily understandable physics.

The basic framework will use a dispersive representation for the
scalar-isoscalar interaction. The overall strength will be governed
by a contact interaction
\begin{equation}
H_{contact} = G_S \bar{N}N\bar{N}N +...
\end{equation}
The strength of this interaction will be related to a dispersion
relation\cite{vinhmau, durso}
\begin{equation}
G_{S} = \frac{2}{\pi}\int_{2m_\pi}^{\infty} \frac{d\mu}{\mu}
~{\rho}_{S}(\mu^2)
\end{equation}
Here the spectral function in the integrand $\rho_S(\mu)$ is to be
calculated from physical intermediate states of two-pion exchange,
using chiral amplitudes at low energy plus the Omnes function for
pion rescattering. We can control the mass dependence of these
ingredients to a great extent using chiral perturbation theory.

\begin{figure}[ht]
 \begin{center}
  \begin{minipage}[t]{.07\textwidth}
    \vspace{0pt}
    \centering
    \vspace*{0 in}
    \hspace*{70pt}{$\rho(\mu)/\mu$} \\
  \end{minipage}%
  \begin{minipage}[t]{0.93\textwidth}
    \vspace{0 in}
    \centering
\hspace{-0.0 in}\rotatebox{-0}
{\includegraphics[width=0.6\textwidth,height=!]{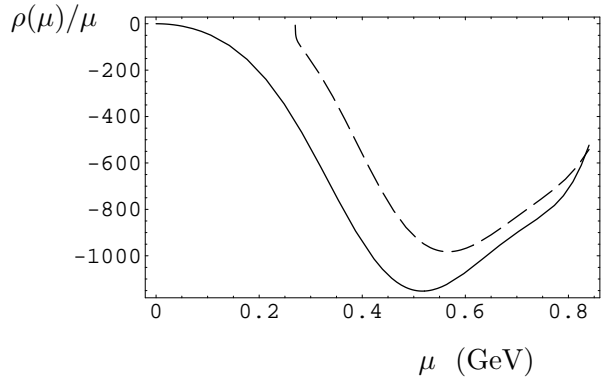}}\\
    ${} \vspace*{-0pt}$
\mbox{\hspace*{100pt} {$\mu$} ~(GeV)}\\
  \end{minipage}
 \end{center}
 \caption{The spectral integrand, $\rho_S(\mu)/\mu$, whose integral determines the strength of the
 scalar-isoscalar interaction. The curve that extends down to $\mu=0$ (solid line) is that for
 the chiral limit - the other (dashed line) is calculated with the physical pion mass.}
 \label{spectralresults}
\end{figure}
Let us at this stage show the basic result, to be developed more
fully below. The spectral function for the physical case was
developed in Ref. \cite{sigma} where it was shown to reproduce the
shape and magnitude expected from past experience with scalar
exchange. This result, with a small modification to be described
below, is shown in Fig. 1. Also shown is the result that will be
developed for the spectral function in the chiral limit. The
integral under theses curves gives the strength of the scalar
interaction\footnote{The negative ``peak'' in this figure is
phenomenologically equivalent to the exchange of a ``$\sigma$''
resonance}. The chiral limit is seen to have a greater strength
largely because the threshold extends down to zero energy and the
spectral function develops quickly. As will be discussed more fully
below, there is some modeling involved in producing these spectral
functions, and there are some gaps in our understanding of mass
effects. However, the bulk of the change in the chiral limit comes
from relatively low energies, where we have better control over the
calculation. These low energy effects are easy to defend as
predictions of chiral perturbation theory.

These contact interactions enter modern descriptions of nuclear
binding. They have been developed in a systematic fashion in the
treatment of light nuclei. Below I will also discuss the
modification in the binding energy in heavy nuclei using the contact
interactions. We will find a large increase in the binding energy
per nucleon.

 In chiral perturbation theory, the coupling constants
will have an expansion in the pion mass. The leading terms will be
\begin{equation}
G_S = G_{S0}(1 + d_S m_\pi^2) + F_S m_\pi^3 +...
\end{equation}
Equivalently one sees also the parameter $D_S \equiv d_S G_S$
defined. I will use my results to generate a value for the chiral
slope, $d_S$.

The results indicate rather large shifts as one goes to the chiral
limit. In particular, I will find
\begin{eqnarray}
\eta_{S,~ch} &=& \frac{G_S|_{chiral}}{G_S|_{physical}} = 1.37
\pm 0.10  \nonumber \\
d_S &=& \frac{0.31 \pm 0.08}{m_\pi^2}
\end{eqnarray}
When unexpectedly large effects are found, it is important to
carefully understand their origin and assess whether the effects are
reasonable. I spend considerable effort on this task in this paper
and we can in fact understand why a large effect is found.

In section 2, I describe some properties of effective field theory
that make it reasonable that such a large mass variation exists. In
Section 3, the basic framework is displayed. Section 4 discusses the
leading chiral amplitudes for the spectral function. Section 5
describes the Omnes function and the the framework that I use for
approximating the phase shifts. Section 6 puts these ingredients
together and shows the comparison of the physical case to the chiral
limit. I also discuss the uncertainties of the calculation here.
Section 7 uses this scalar coupling to describe nuclear binding in
the chiral limit. The slope away from the chiral limit $d_S$ is
extracted in Section 8. Finally in section 9 I briefly summarize the
results and discuss the relation of this work to previous treatments
of the nuclear force in the chiral limit\cite{meissnerchiral,
savagechiral,lattice}.

\section{Effective field theory and low energy effects}

There is a dichotomy between the way that we describe the philosophy
of effective field theory and and the way that we implement it in
practice. The difference is important for the understanding of the
present problem. Let me introduce these notions through a simpler
problem which is by now well understood, and which also has an
anomalously large mass dependence - namely the pion and kaon
electromagnetic mass splitting. I then show how this philosophy is
relevant to the problem of the scalar interaction in nuclei.

An effective field theory is the description of the low energy limit
of a theory, with the correct degrees of freedom and interactions
for the low energy world. As originally explained by
Wilson\cite{wilson}, we describe the low energy interaction below
some scale $\Lambda$ by the propagation of the light degrees of
freedom. The high energy effects above the scale $\Lambda$ are
incorporated into a set of local Lagrangians. They must be local at
energies $E<< \Lambda$ due to the uncertainty principle. We then
have a clear way of separating low energy effects, which we can
readily calculate, from high energy effects, which we often cannot
calculate but which can readily parameterized.

However, this is not how we generally proceed in practice. The use
of a cutoff  $\Lambda$ to separate low energy from high energy is
awkward. Energy cutoffs require care in order to not violate Lorentz
invariance, gauge invariance and/or chiral symmetry. In practice we
generally calculate using dimensional regularization. However, this
scheme does not have a separation between low energy and high energy
- all scales are integrated over in loop diagrams. This is not a
problem in principle. Extra or missing contributions can be adjusted
by a shift in the values of the contact interactions. However, and
this will be our topic here, unexpectedly large mass effects can
occur if large {\em low energy} contributions are put into the
contact interaction.

\begin{figure}[h]
 \begin{fmffile}{fmfemloop6}
  \vspace{0.5in}
  \begin{equation*}
   \begin{fmfgraph*}(100,50)
    \fmfleft{vleftlow} 
    \fmfright{vrightlow} 
  \fmf{plain}{vleftlow,vmiddlelow1,vmiddlelow2,vmiddlelow3,vrightlow}\
  \fmfpen{thick}
  \fmf{dbl_plain}{vmiddlelow1,vmiddlelow2,vmiddlelow3}
  \fmfblob{0.07w}{vmiddlelow1,vmiddlelow3}
 \fmfv{label=${\pi/K}$,label.angle=180}{vleftlow}
\fmfv{label=${\pi/K}$,label.angle=0}{vrightlow}
    \fmffreeze
    \fmf{photon,left,tension=0,label=$\gamma$,label.side=left}{vmiddlelow1,vmiddlelow3}
   \end{fmfgraph*}  {}\hspace*{30pt}
  \end{equation*}
  \end{fmffile}
$\hspace*{0.1\textwidth}$ \vspace{-0.2 in} \caption{The photon loop
diagram which contributes to the pion and kaon electromagnetic mass
differences.} \vspace*{10pt}
  \end{figure}
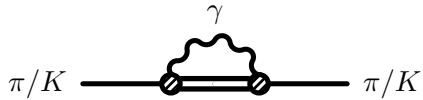

 Let us explore this first in the context of the pion
electromagnetic mass difference. This is described by the chiral
Lagrangian
\begin{equation}
{\cal L}_{EM} = g_{EM} ~Tr(QUQU^\dagger)
\end{equation}
in usual notation. One consequence of this is the equality of the
electromagnetic mass splitting of pions and kaons
\begin{equation}
(m^2_{\pi^+} -m^2_{\pi^0})_{EM} = (m^2_{K^+}-m^2_{K^0})_{EM}
\end{equation}
which is known as Dashen's theorem\cite{Dashen}. Normally
corrections to this would be expected to be of order 25\% from the
naive dimensional analysis estimate of SU(3) breaking due to quark
mass differences. However direct calculation of the mass differences
in chiral-based models yields a 100\% violation of Dashen's
theorem\cite{emmodels}. The reason why the models are right and the
naive dimensional analysis estimate is wrong has to do with the
distinction described above.

In the effective field theory the degrees of freedom are pions,
kaons and photons. In a Wilsonian effective field theory, the
quantum effects of the light particles in the diagram of Figure 2
should be calculated up to the scale $\Lambda$, and only the effects
beyond that scale parameterized by the chiral Lagrangian with
coefficient $g_{EM}(\Lambda)$. Thus in a Wilsonian scheme there
would be two different contributions: a dynamical contribution of
the actual $\pi, ~K, ~\gamma$ loop diagrams up to the scale
$\Lambda$ and a parameter $g_{EM}(\Lambda)$ describing the physics
beyond the scale $\Lambda$. There will also be higher order terms in
the Lagrangian with extra factors of the pion and kaon masses.
While we expect only a modest variation of the chiral Lagrangian
with quark mass - this is the real content of Dashen's theorem -
there is no such guarantee for dynamical effects of the light
particles.

In dimensional regularization the structure is different. The photon
loop in dimensional regularization has the form (for the pion self
energy)
\begin{equation}
\sim \frac{e^2}{16\pi^2}\frac{m_\pi^2}{d-4}
\end{equation}
The factor of $m_\pi^2$ is forced by dimensional analysis - there is
no other dimensional factor in the calculation. Because of this
factor, the loop does not renormalize $g_{EM}$, but goes into the
renormalization of a higher dimension term in the lagrangian that
has extra factors of $m_\pi^2$ and $m_K^2$ - let us call this
parameter $g_{EM,~m^2}$. So in dimensional regularization there is
no residual dynamical loop contribution, and the analysis only
involves the parameters of the chiral Lagrangian treated at tree
level.

\begin{figure}[ht]
 \begin{center}
   \begin{minipage}[t]{.07\textwidth}
    \vspace{0pt}
    \centering
    \vspace*{0 in}
    \hspace*{40pt}{$-Q^2\Pi_{V-A}(Q^2)$}
  \end{minipage}%
  \begin{minipage}[t]{0.93\textwidth}
    \vspace{0 in}
    \centering
\hspace{-0.0 in}\rotatebox{-0}
{\includegraphics[width=0.6\textwidth,height=!]{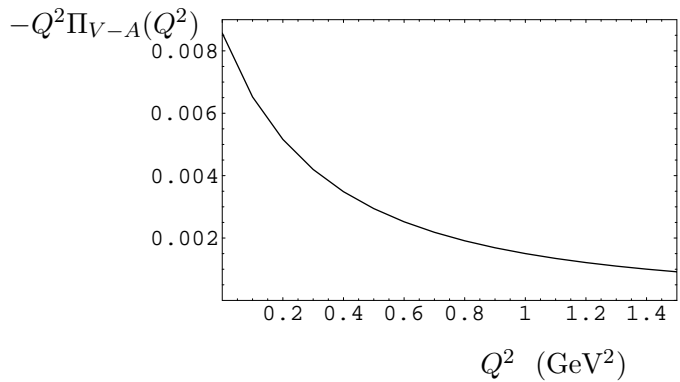}}\\
    ${} \vspace*{-05pt}$
    \mbox{\hspace*{100pt} {$Q^2$} ~(GeV$^{2}$)} \\
  \end{minipage}
 \end{center}
 \caption{The figure shows the vacuum polarization
 function $Q^2\Pi_{V-A}(Q^2)$ whose integral determines the chiral parameter $g_{EM}$.
 Notice that the greatest contribution comes from low energy.}
 \label{em}
\end{figure}

It is easy to demonstrate that the greatest contributions to the
chiral coefficient $g_{EM}$ comes from low energy physics, not from
high energy. There is a dispersive sum rule \cite{das} for $g_{EM}$,
or equivalently the pion electromagnetic mass difference in the
chiral limit, for which data exists. In the chiral limit, we have
\begin{equation}
g_{EM} = -\frac{3\alpha}{8\pi}\int_0^\infty dQ^2 ~Q^2\Pi_{V-A}(Q^2)
\end{equation}
where $\Pi_{V-A}$ is the difference of the vector and axial-vector
vacuum polarization functions\footnote{There is an extensive
literature on $\Pi_{V-A}$ - the figure is drawn from my own work on
the subject\cite{cirigliano}.}. In turn there is a spectral sum rule
for $\Pi_{V-A}$, and the input to this sum rule can be obtained from
$\tau$ decay data. The integrand, derived from ALEPH data, is shown
if Fig. \ref{em}. We see directly that most of the input comes from
low energy.

Now consider what happens if we adopt the Wilsonian scheme and the
dynamical low energy contribution is relatively large. This
dynamical effect can easily (and does) have over a 100\% variation
when comparing the pion with the kaon, because $m_K^2 \sim 13
m_\pi^2$ and we are in an energy region comparable to the masses. In
this case, even if the mass dependence of short distance physics
parameterized by the chiral Lagrangian is of normal size, the kaon
mass shift can be very much different from that of the pion.

The Wilsonian approach is undoubtedly correct in saying that there
is large SU(3) breaking at low energy. It is readily calculable
using just the low energy theory. However, when doing chiral
perturbation theory in the usual fashion, regularized dimensionally,
we would miss this effect. Dimensional regularization can
accommodate this situation, but it is not the outcome that we would
naively expect. It can be done only at the expense of having an
anomalously large parameter in the chiral Lagrangian. The mass shift
is described by $g_{EM}$ and $g_{EM,~m^2}$, which are a priori
unknown. The latter must be taken to be much larger than naive
expectation in order to reproduce the large difference.

What has happened is that, in this scheme, important {\em low energy
physics} has been encoded in the parameter of the chiral Lagrangian
$g_{EM,~m^2}$. We normally expect that the chiral parameters are
manifestations of only high scale physics but in a scheme like
dimensional regularization, that does not have a separation between
high and low scales, there can also be low energy physics encoded in
the parameters.

The lesson from this is that if low energy dynamical effects make a
significant contribution to a process there can be a large mass
effect. This can lead to a larger-than-expected variation in chiral
parameters when treated in the usual fashion. I will now argue that
an analogous effect will occur in the parameter $G_S$ describing the
strength of the scalar interaction. That is, we will see that low
energy effects are embedded in this parameter and that these effects
can have a very large variation.

Consider now the two pion exchange contributions to the internucleon
potential. The diagrams and formulas will be given later. In a
Wilsonian approach we would calculate explicitly the pionic effects
up to some separation scale $\Lambda$, and then use a contact
interaction $G_S(\Lambda)$ to describe the high energy effects
beyond this scale. We might realistically expect that the mass
effects in $G_S(\Lambda)$ are modest. However, there can be no such
expectations for the explicit two pion calculation if there are
significant contributions from energies around the pion mass. These
contributions can vary by 100\% as the pion mass is set to zero. For
example, in Figure 1 we see that the physical spectral function can
only start at $\mu = 2 m_\pi$ while the chiral limit case extends
down to zero energy. The two cases do not get close to each other
until above 400 MeV, and the integrated values of the contribution
below this energy are significantly different.

However this Wilsonian approach is  usually not carried out in
practice. The interaction in the scalar channel is parameterized by
a contact interactions (or a sigma potential) and the low energy
parts of the two pion exchange are not separately evaluated. One
calculates nuclear binding totally with the contact interaction or
the potential. The consequence of this is that we should expect that
the large mass variation of the dynamical two-pion exchange has to
be included in the contact interaction, which then appears as if
there were an anomalously large mass dependence of this coefficient.
We see in Figure 1 that the high energy portion has a modest mass
dependence while the low energy component has a 100\% variations
between the physical case and the chiral limit. If we were following
the Wilsonian approach, the parameter $G_S(\Lambda)$ would have a
small mass variation if $\Lambda$ were chosen above 500 MeV.
However, in a usual calculation the mass variation of $G_S$, encoded
in the parameter $d_S$, must be taken to be very large in order to
account for the low energy effect. It is this which allows us to
understand the large mass effect described in the following
sections.

It should be said that there is now an attempt at what I call the
Wilsonian approach. Epelbaum, Gl\"{o}ckle and Mei{\ss}ner\cite{egm,
meissnerrev} have been calculating the pionic effects to the nuclear
potential in chiral perturbation theory using a cutoff in the
spectral integral. When supplemented with a cutoff dependent contact
interactions this is the procedure described above. It would be good
to see this approach extended to the binding of heavy nuclei. It
would be even preferable, without loss of generality, to use the
Omnes formalism developed in the present paper for the low energy
amplitude, even if a cutoff is used. The Omnes function is a
required feature of the full answer and tames the growing polynomial
behavior of the chiral amplitudes such that there is not an
excessive contribution from the energy region around the cutoff.
This procedure is equally general because any mistake that is
introduced from the region around the cutoff can nevertheless be
corrected for by an adjustment of the contact term.

\section{Formalism}

Our procedure begins with the dispersion relation derived by
Cottingham, Vinh Mau and others\cite{vinhmau, durso}. For textbook
reviews, see \cite{brown, ericson}. The scattering amplitude for two
nucleons obeys an unsubtracted t-channel dispersion relation.
\begin{equation}
 M(s,t) = \frac{1}{\pi}\int_{4m_\pi^2}^\infty d\mu^2~\frac{Im M(s,\mu^2)}{\mu^2
-t -i\epsilon }
\end{equation}
The imaginary part of this amplitude is connected to the crossed
channel $N\bar{N}\rightarrow N\bar{N}$ with the important
intermediate state being that of two pions. The overall amplitude is
decomposed into partial waves described by their spin and isospin
quantum numbers. The greatest interest in this paper will be on the
scalar-isoscalar (J=0, I=0) channel. By taking the nonrelativistic
limit and ignoring the energy dependence in the S channel, one can
define a momentum space potential that depends only on the momentum
transfer $q^2$. Using the scalar isoscalar central potential as an
example, let us define the corresponding spectral function
\begin{equation}
\rho_S (\mu) = Im V_S (q = i \mu)
\end{equation}
In terms of this imaginary part the potential is defined by the
dispersion relation
\begin{equation}
{V}_S(q^2) =  \frac{2}{\pi}\int_{2m_\pi}^\infty d\mu~\mu
\frac{\rho_S(\mu)}{\mu^2+q^2}
\end{equation}
When converted coordinate space this defines an inter-nucleon
potential depending on the separation $r$, with a spectral
representation
\begin{equation}
V_S(r) =  \frac{1}{2 \pi^2 r}\int_{2m_\pi}^\infty d\mu~\mu ~e^{-\mu
r}~{\rho}_S(\mu)
\end{equation}
For orientation, note that this description would produce the
conventional sigma exchange potential with the substitution
\begin{equation}
\rho_S(\mu) = -\pi g_\sigma^2 \delta(\mu^2 - m_\sigma^2)
\end{equation}
This would recover the classic Yukawa potential in momentum and
coordinate space:
\begin{equation}
V_\sigma =\frac{-g^2_\sigma}{q^2+m^2_\sigma -i\epsilon}
\end{equation}
and
\begin{equation}
V_S(r) = - \frac{g_\sigma^2}{4 \pi r} ~e^{-m_\sigma r}
\end{equation}
However, the $\delta$-function is not a good representation of the
physics in the scalar channel. Instead we will use chiral
perturbation theory extended to higher energy with the Omnes
representation to describe the ingredients to the scalar channel.

For all channels except that of one-pion-exchange, the potential is
short range, and in effective field theory can be represented by a
delta function - a contact interaction. For example, the scalar and
vector isoscalar contact interactions would be represented as
\begin{equation}
H_{contact} = G_S \bar{N}N\bar{N}N + G_V \bar{N}\gamma_\mu
N\bar{N}\gamma^\mu N +...
\end{equation}
Since the Fourier transform of a constant is a delta function, the
contact interactions in coordinate space correspond to a constant in
momentum space. These contact interactions are therefore given by
the momentum space potential at $q^2=0$, i.e. for the scalar or
vector channel
\begin{equation}
G_{S,V} = V_{S,V}(q^2=0)=
\frac{2}{\pi}\int_{(2m_\pi,3m_\pi)}^{\infty} \frac{d\mu}{\mu}
~{\rho}_{S,V}(\mu^2)
\end{equation}
Higher powers of momenta can be accommodated by derivative contact
interactions. For the exchange of a narrow resonance, the contact
interaction has the form $G_{i} = \pm g^2/m_i^2$, and I will use
this relation for the vector meson effect\footnote{For resonance
saturation estimates of these and other nuclear contact
interactions, see \cite{resonance}.} However, for the scalar
interaction it is important to have a more complete evaluation, as
described below.

\section{Chiral amplitudes}

The application of chiral perturbation theory to the nuclear force
has an extensive literature - for reviews and references, see
\cite{meissnerrev, beanerev, bedaquerev, epelbaumrev}. The pathway
using the dispersion relation and the spectral function has been
pioneered by Kaiser, Mei{\ss}ner and collaborators\cite{brockmann,
Kaiser, morekaiser, egm}. We will find this a very effective
description of the nature of the energy expansion for the nuclear
interaction.

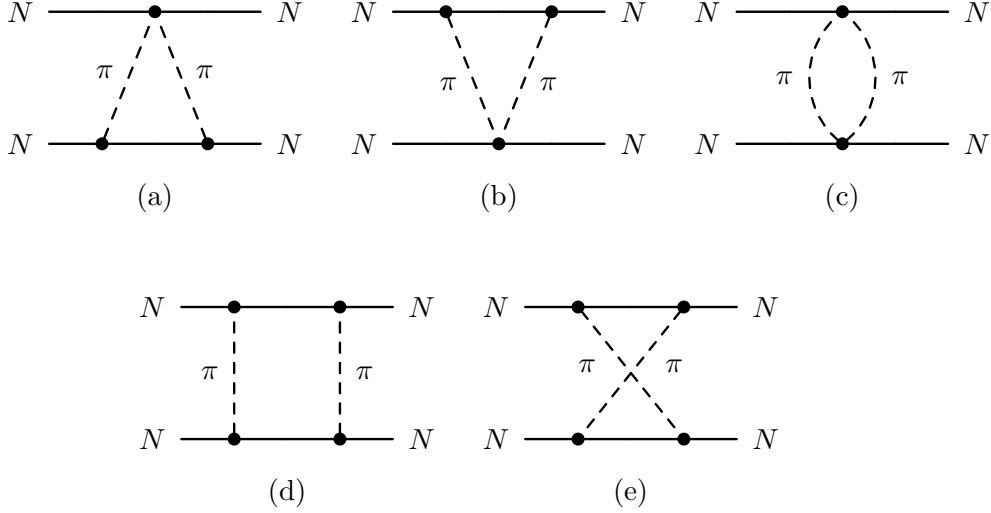
\begin{figure}[h]
 \begin{fmffile}{fmftwopion}
  \begin{equation*}
   \begin{fmfgraph*}(100,50)
    \fmfleft{vleftlow,vlefthigh} 
    \fmfright{vrightlow,vrighthigh} 
    \fmf{plain}{vleftlow,vmiddlelow1,vmiddlelow2,vmiddlelow3,vrightlow} 
        \fmf{plain}{vlefthigh,vmiddlehigh1,vmiddlehigh2,vmiddlehigh3,vrighthigh} 
    \fmfdot{vmiddlelow1,vmiddlelow3,vmiddlehigh2}
\fmfv{label=$N$,label.angle=180}{vlefthigh}
\fmfv{label=$N$,label.angle=180}{vleftlow}
\fmfv{label=$N$,label.angle=0}{vrighthigh}
\fmfv{label=$N$,label.angle=0}{vrightlow}
    \fmffreeze
    \fmf{dashes,label=$\pi$,label.side=left}{vmiddlelow1,vmiddlehigh2}
        \fmf{dashes,label=$\pi$,label.side=right}{vmiddlelow3,vmiddlehigh2}
        \fmfv{label={(a)},label.angle=-90,label.dist=15}{vmiddlelow2}
   \end{fmfgraph*}  {}\hspace*{30pt}
      \begin{fmfgraph*}(100,50)
    \fmfleft{vleftlow,vlefthigh} 
    \fmfright{vrightlow,vrighthigh} 
    \fmf{plain}{vleftlow,vmiddlelow1,vmiddlelow2,vmiddlelow3,vrightlow} 
        \fmf{plain}{vlefthigh,vmiddlehigh1,vmiddlehigh2,vmiddlehigh3,vrighthigh} 
    \fmfdot{vmiddlelow2,vmiddlehigh3,vmiddlehigh1}
\fmfv{label=$N$,label.angle=180}{vlefthigh}
\fmfv{label=$N$,label.angle=180}{vleftlow}
\fmfv{label=$N$,label.angle=0}{vrighthigh}
\fmfv{label=$N$,label.angle=0}{vrightlow}
    \fmffreeze
    \fmf{dashes,label=$\pi$,label.side=right}{vmiddlehigh1,vmiddlelow2}
        \fmf{dashes,label=$\pi$,label.side=left}{vmiddlehigh3,vmiddlelow2}
        \fmfv{label={(b)},label.angle=-90,label.dist=15}{vmiddlelow2}
   \end{fmfgraph*}  {}\hspace*{30pt}
   \begin{fmfgraph*}(100,50)
    \fmfleft{vleftlow,vlefthigh} 
    \fmfright{vrightlow,vrighthigh} 
    \fmf{plain}{vleftlow,vmiddlelow1,vmiddlelow2,vmiddlelow3,vrightlow} 
        \fmf{plain}{vlefthigh,vmiddlehigh1,vmiddlehigh2,vmiddlehigh3,vrighthigh} 
    \fmfdot{vmiddlelow2,vmiddlehigh2}
\fmfv{label=$N$,label.angle=180}{vlefthigh}
\fmfv{label=$N$,label.angle=180}{vleftlow}
\fmfv{label=$N$,label.angle=0}{vrighthigh}
\fmfv{label=$N$,label.angle=0}{vrightlow}
    \fmffreeze
  \fmf{dashes,left=0.5,label=$\pi$,label.side=left}{vmiddlelow2,vmiddlehigh2}
    \fmf{dashes,right=0.5,label=$\pi$,label.side=right}{vmiddlelow2,vmiddlehigh2}
      \fmfv{label={(c)},label.angle=-90,label.dist=15}{vmiddlelow2}
   \end{fmfgraph*}  {}\hspace*{30pt}
  \end{equation*}
  \vspace{0.5in}
    \begin{equation*}
   \begin{fmfgraph*}(100,50)
    \fmfleft{vleftlow,vlefthigh} 
    \fmfright{vrightlow,vrighthigh} 
    \fmf{plain}{vleftlow,vmiddlelow1,vmiddlelow2,vmiddlelow3,vrightlow} 
        \fmf{plain}{vlefthigh,vmiddlehigh1,vmiddlehigh2,vmiddlehigh3,vrighthigh} 
    \fmfdot{vmiddlelow1,vmiddlelow3,vmiddlehigh1,vmiddlehigh3}
\fmfv{label=$N$,label.angle=180}{vlefthigh}
\fmfv{label=$N$,label.angle=180}{vleftlow}
\fmfv{label=$N$,label.angle=0}{vrighthigh}
\fmfv{label=$N$,label.angle=0}{vrightlow}
    \fmffreeze
    \fmf{dashes,label=$\pi$,label.side=left}{vmiddlelow1,vmiddlehigh1}
        \fmf{dashes,label=$\pi$,label.side=right}{vmiddlelow3,vmiddlehigh3}
        \fmfv{label={(d)},label.angle=-90,label.dist=15}{vmiddlelow2}
   \end{fmfgraph*}  {}\hspace*{30pt}
      \begin{fmfgraph*}(100,50)
    \fmfleft{vleftlow,vlefthigh} 
    \fmfright{vrightlow,vrighthigh} 
    \fmf{plain}{vleftlow,vmiddlelow1,vmiddlelow2,vmiddlelow3,vrightlow} 
        \fmf{plain}{vlefthigh,vmiddlehigh1,vmiddlehigh2,vmiddlehigh3,vrighthigh} 
    \fmfdot{vmiddlelow1,vmiddlelow3,vmiddlehigh3,vmiddlehigh1}
\fmfv{label=$N$,label.angle=180}{vlefthigh}
\fmfv{label=$N$,label.angle=180}{vleftlow}
\fmfv{label=$N$,label.angle=0}{vrighthigh}
\fmfv{label=$N$,label.angle=0}{vrightlow}
    \fmffreeze
    \fmf{dashes}{vmiddlehigh1,vmiddlelow3}
        \fmf{dashes}{vmiddlehigh3,vmiddlelow1}
\fmfv{label=$\pi$,label.side=left,label.dist=20,label.angle=-80}{vmiddlehigh1}
\fmfv{label=$\pi$,label.side=left,label.dist=20,label.angle=-100}{vmiddlehigh3}
        \fmfv{label={(e)},label.angle=-90,label.dist=15}{vmiddlelow2}
   \end{fmfgraph*}  {}\hspace*{30pt}
  \end{equation*}
 \end{fmffile}
$\hspace*{0.515\textwidth}$ \vspace{0 in} \caption{Two pion exchange
diagrams which arise in chiral perturbation theory.} \vspace*{10pt}
\end{figure}

The low energy behavior of the two pion exchange diagrams have been
calculated in perturbation theory. The $\pi NN $ vertex is
proportional to the axial charge $g_A$ and the two pion vertices are
parameterized by low energy constants $c_1,~c_2,~c_3,~c_4$ in the
chiral Lagrangian. The diagrams are shown in Fig. 3. The diagrams of
Fig 3a,b lead to the spectral
function\cite{brockmann}\footnote{Watch out for a sign difference -
their momentum space potential has the opposite sign of the
conventions used in the present paper.}
 \begin{eqnarray}
 \rho_S^{\rm a,b} (\mu) &=& \frac{3 g_A^2}{64
F_\pi^4}\left[4c_1 m_\pi^2 +c_3(\mu^2-2m_\pi^2)
\right]\frac{(\mu^2-2m_\pi^2)}{\mu} ~\theta (\mu -2m_\pi)\nonumber \\
  &&~~~\times\frac{4m_N}{\pi\sqrt{4m_N^2 -
\mu^2}}\arctan\frac{\sqrt{(\mu^2
-4m_\pi^2)(4m_N-\mu^2)}}{\mu^2-2m_\pi^2}
 \label{rho1}
\end{eqnarray}
The result of \cite{brockmann} was calculated in the heavy baryon
limit and I have modified their result to include the effects of a
finite nucleon mass \cite{becher}, see also\cite{higa}. The
difference is minor numerically. Diagram 3b is one power higher in
the energy expansion and involves two factors of the low energy
constants $c_{1,2,3}$ The imaginary part of this diagram is
\cite{Kaiser}
\begin{eqnarray}
\rho_S^{\rm c} (\mu) &=& -\frac{3}{32 \pi
F_\pi^4}\sqrt{1-\frac{4m_\pi^2}{\mu^2}} ~\theta (\mu -2m_\pi)  \\
& &~ \left(\left[4c_1 m_\pi^2 +\frac{c_2}{6}(\mu^2 - 4m_\pi^2)
+c_3(\mu^2-2m_\pi^2) \right]^2 +\frac{c_2^2}{45}(\mu^2-4
m_\pi^2)^2\right)\nonumber \label{rho2}
\end{eqnarray}

Figures 3d,e, the box and cross box diagrams, have an ambiguous
status in the calculation of the contact interaction. In an
effective field theory treatment, one includes the $\pi NN$ vertex
in the low energy Lagrangian and treats pion exchange dynamically,
including loop processes. This procedure would generate these two
diagrams explicitly. Therefore in principle, the correct treatment
is to treat the box and cross box separately from the contact
interaction. In practice is is not always clear how much of these
diagrams are calculated dynamically and how much is parameterized.
This probably accounts for the relatively wide range of values of
$G_S$ that one can find in the literature. In a relativistic
treatment such as applied in Walecka-style models which I use in
Sec. 8, one would expect that these terms should not be included in
the contact term. This is because the pion and nucleon propagators
are treated relativistically, in which case the box and crossed box
diagrams would be calculated directly and should not be double
counted by also appearing in the contact term. For this reason we
will not include them in our calculation of the contact interaction.
We are also fortunate that, on a purely practical level, these terms
are small compared to the effects that we do study, at least for the
scalar central potential. This is demonstrated in \cite{brockmann}
and and can also be seen in Fig 3.15 of Ref. \cite{ericson}. As an
explicit example, if take the $q^2 =0$ limit of the potential from
these diagrams quoted in \cite{brockmann}, one finds that they
generate a shift in the non-analytic mass correction of size
\begin{equation}
\Delta G_S = -\frac{15g_A^4m_\pi^3}{1024\pi F_\pi^4 M}= 0.42 ~{\rm
GeV}^{-2}
\end{equation}
where the total result is $G_S\sim-400~$GeV$^{-2}$. For these
reasons we will not use Fig 3d,e in our calculation of the contact
interaction.

\section{Unitarity and the Omnes function}

A key ingredient in the present method is the use of the Omnes
function, which incorporates the physical effects of pion
rescattering\cite{Omnes}. This ingredient is not optional - it is
required by unitarity. Moreover the general form of the result is
well know within the elastic region. The amplitude must have the
form of a polynomial in the energy times the Omnes function
\begin{equation}
\Omega (\mu) = exp[\frac{\mu^2}{\pi} \int
\frac{ds}{s}~\frac{\delta(s)}{s-\mu^2}]
\end{equation}
Here $\delta$ is the $\pi\pi$ scattering phase shift, in our case
for the I=0, J=0 channel. Chiral perturbation theory is consistent
with this order by order in the energy expansion. Following Ref.
\cite{dgl}, it is known how to match this general description to the
results of chiral perturbation theory by appropriately identifying
the polynomial. The two pion vertices that appear in Fig. 3a,b,c are
modified by this Omnes function in a way that is strictly analogous
to the two pion vertex described in\cite{dgl}.

Experience has shown that using the lowest order amplitudes
supplemented by the Omnes function gives reasonably good results up
to 700-800 MeV. I have elsewhere \cite{sigma} applied this formalism
to the nuclear central potential using the physical phase shifts
matched to the lowest order results in order to generate the
spectral function. The result is very encouraging because it
generates the main features known phenomenologically for this
interaction. It was shown that this give good results for the shape
and magnitude of the spectral function in the scalar channel. In
traditional treatments this shape is described by a broad low mass
``$\sigma$'' particle, although it is unlikely that such a resonance
is present in the spectrum of QCD. In particular, in the treatment
of \cite{sigma} the spectral function is given by
\begin{equation}
\rho_S(\mu) = \rho_S^{a,b} Re \Omega(\mu) +
\rho_S^{c}|\Omega(\mu)|^2 \label{spectral}
\end{equation}
The effect of the Omnes function is to reshape the spectral function
so that it mimics a $\sigma$ even though such a resonance is not
present in reality. I will continue to use this formalism in the
present paper. While the use of the Omnes function is rigorous and
can be matched to chiral perturbation theory to any given order, the
use of only the lowest order amplitudes in this expression is
clearly not rigorous except at low energies - it is here that the
modeling of the spectral function enters.

We have recorded the chiral amplitudes in the previous section. The
task now is to understand how these amplitudes vary between the
physical case and the chiral limit. There will be several
ingredients that need to be explored. Some mass dependence is
explicit in the chiral amplitudes. Other dependence is contained in
the parameters $F_\pi~g_A$. Finally there is mass dependence in the
Omnes function, or equivalently in the $\pi\pi$ phase shifts. This
latter feature requires the most work, and I will address it first.

At low momentum, we have explicit expressions for the $\pi\pi$
scattering amplitudes through the work of Gasser and
Leutwyler\cite{gasser}. At higher energies, our best knowledge comes
from a treatment that combines chiral symmetry, dispersion
relations, crossing and experimental data\cite{cgl} by Colangelo,
Gasser and Leutwyler (CGL). In \cite{sigma} I used the CGL phase
shifts to construct the Omnes function. However, because of the
reliance on experimental data, we don't directly have the ability to
vary the quark mass in this analysis. In order to calculate the
Omnes function for the chiral limit, we need to introduce a method
to model the high energy behavior of the $\pi\pi$ phase shifts in
such a way that we can vary the pion mass.

\begin{figure}[ht]
 \begin{center}
  \begin{minipage}[t]{.07\textwidth}
    \vspace{0pt}
    \centering
    \vspace*{0 in}
    \hspace*{80pt}{$\delta$} \\
  \end{minipage}%
  \begin{minipage}[t]{0.93\textwidth}
    \vspace{0 in}
    \centering
\hspace{-0.0 in}\rotatebox{-0}
{\includegraphics[width=0.6\textwidth,height=!]{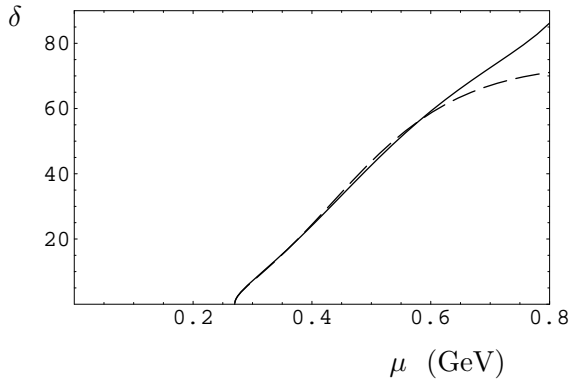}}\\
    ${} \vspace*{-0pt}$
\mbox{\hspace*{100pt} {$\mu$} ~(GeV)}\\
  \end{minipage}
 \end{center}
 \caption{The $\pi\pi$ phase shift from CGL (solid line) and that
 found from the inverse amplitude method (dashed line). The agreement is excellent up
 to about 700 MeV.}
 \label{phasecgl}
\end{figure}

There exists a good and successful method for extending the
description of the chiral amplitudes, often referred to as Pad\'{e}
approximation or the inverse amplitude method(IAM)\cite{iam}. This
provides an analytic approximation to the scattering amplitudes in a
form that is fully satisfies unitary and which can be matched order
by order to the results of chiral perturbation theory. Matched to
the results at order $E^4$, one has the result for the $I=0,~ J=0$
amplitude is
\begin{equation}
T_{00} = t_2 + t_4 +...\rightarrow {t_2^2 \over t_2 - t_4}
\end{equation}
The second term $t_4$ contains the effects of loop diagrams and
hence also has imaginary parts. The beauty of the IAM representation
is that this simple rearrangement allows the amplitude to satisfy
unitarity exactly. While the method necessarily differs from the
full answer beyond the order to which has been matched\footnote{For
example some two-loop logarithms are missing when the IAM is matched
at one loop order\cite{nextorder}}, studies have shown that this
representation provides a good description of the scattering
amplitudes up to reasonably large energies. For example, the
scalar-isoscalar IAM amplitude is compared to the physical result of
CGL in Fig \ref{phasecgl}.
The description is quite good up to 700 MeV, after which it falls
short. The resulting Omnes function is also compared to that derived
from the CGL phase shifts in Fig 5. \footnote{In producing the Omnes
function, I had to extend the phase shifts above the $\mu$ = 850 MeV
endpoint of the CGL analysis in order that the principle value part
of the Omnes function integral be well behaved near the upper end.
Likewise in the IAM formalism, there is a corresponding extension.
As long as this extension is smooth it has little effect on this
calculation. }. Again the results are similar, except at high
energies where the Omnes function is small.

\begin{figure}
\begin{center}
 \begin{minipage}[t]{.07\textwidth}
    \vspace{0pt}
    \hspace*{-180pt}
    \rotatebox{0}{$|\Omega|^2$}
    \hspace*{195pt}{$Re\Omega$}
     \end{minipage}
  \begin{minipage}[t]{0.93\textwidth}
  \vspace{-30pt}
$\begin{array} {c@{\hspace{0.01 in}}c} \multicolumn{1}{l}{} &
\multicolumn{1}{l}{} \\
{\resizebox{2.5in}{!}{\includegraphics{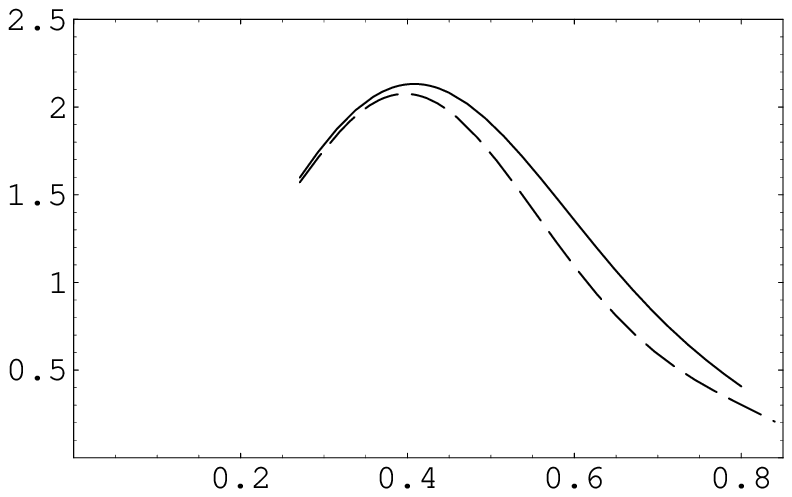}}}&
~~~~~~~~{\resizebox{2.5in}{!}{\includegraphics{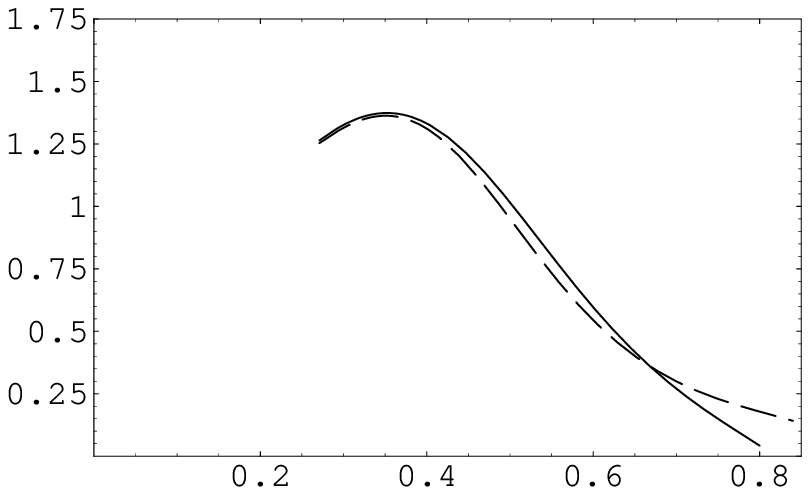}}} \\
\mbox{\hspace*{100pt} {$\mu$} ~(GeV)} & \mbox{\hspace*{75pt} {$\mu$}~(GeV)} \\
\mbox{\hspace*{25pt} (a)} & \mbox{\hspace*{25pt} (b)}
\end{array}$
  \end{minipage}
\end{center}
\caption{The absolute square (a) and the real part (b) of the Omnes
functions derived from the CGL phase shifts (solid line) and those
found using the inverse amplitude method (dashed line). }
\label{cglOmnes}
\end{figure}

I will use the IAM phase shifts as an analytic approximation of the
true amplitudes, keeping in mind the shortcomings at higher
energies. The low energy behavior of these phase shifts will then be
fully rigorous and the high energy portion will be approximate.

\begin{figure}[ht]
 \begin{center}
   \begin{minipage}[t]{.07\textwidth}
    \vspace{0pt}
    \centering
    \vspace*{0 in}
    \hspace*{80pt}{$\delta$} \\
  \end{minipage}%
  \begin{minipage}[t]{0.93\textwidth}
    \vspace{0 in}
    \centering
\hspace{-0.0 in}\rotatebox{-0}
{\includegraphics[width=0.6\textwidth,height=!]{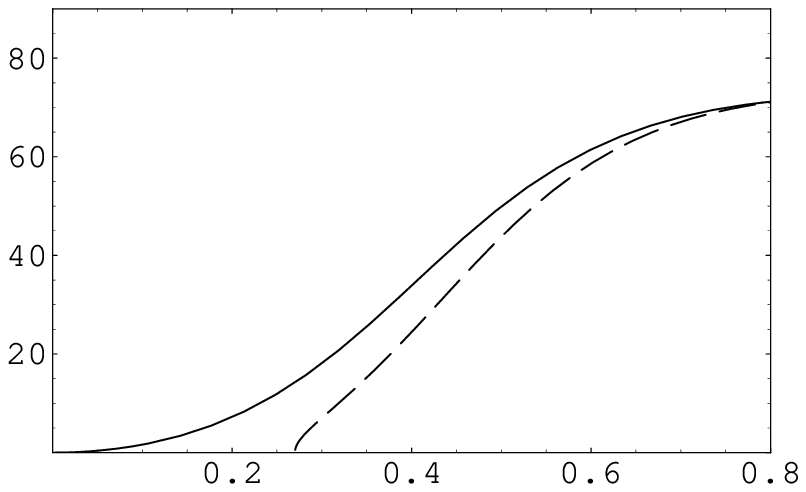}}\\
    ${} \vspace*{-0pt}$
\mbox{\hspace*{100pt} {$\mu$} ~(GeV)}\\
  \end{minipage}
 \end{center}
 \caption{The scalar-isoscalar $\pi\pi$ phase shifts using the IAM in the physical
 case (dashed line) and in the chiral limit (solid line).}
 \label{chiral phases}
\end{figure}

\begin{figure}
\begin{center}
 \begin{minipage}[t]{.07\textwidth}
    \vspace{0pt}
    \hspace*{-180pt}
    \rotatebox{0}{$|\Omega|^2$}
    \hspace*{195pt}{$Re\Omega$}
     \end{minipage}
  \begin{minipage}[t]{0.93\textwidth}
  \vspace{-30pt}
$\begin{array} {c@{\hspace{0.01 in}}c} \multicolumn{1}{l}{} &
\multicolumn{1}{l}{} \\
{\resizebox{2.5in}{!}{\includegraphics{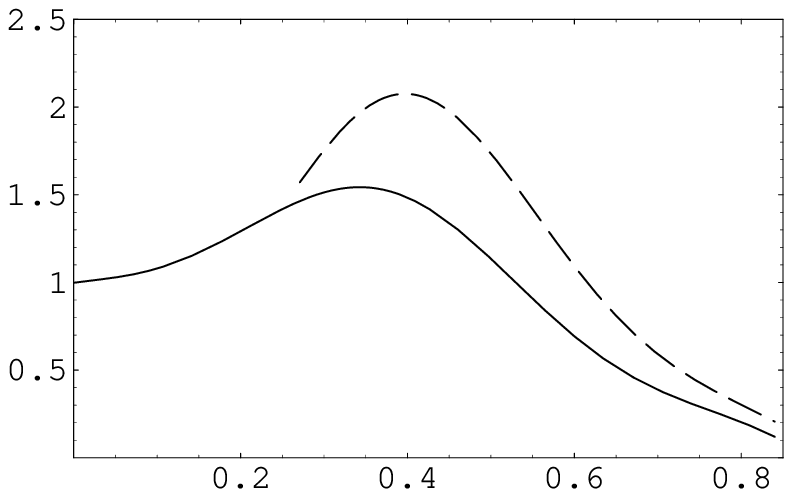}}}&
~~~~~~~~{\resizebox{2.5in}{!}{\includegraphics{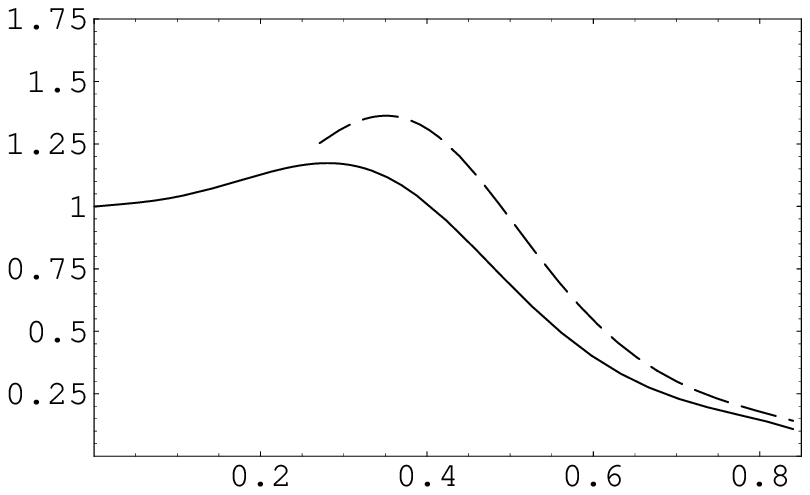}}} \\
\mbox{\hspace*{100pt} {$\mu$} ~(GeV)} & \mbox{\hspace*{75pt} {$\mu$}~(GeV)} \\
\mbox{\hspace*{25pt} (a)} & \mbox{\hspace*{25pt} (b)}
\end{array}$
  \end{minipage}
\end{center}
\caption{The absolute square (a) and the real part (b) of the Omnes
functions in the physical
 case (dashed line) and in the chiral limit (solid line).} \label{chiral Omnes1}
\end{figure}

Let us immediately look at the phase shift and the Omnes function in
the chiral limit. The phase is shown in figure \ref{chiral phases}
in comparison with the IAM approximation to the physical case. The
salient feature is the threshold behavior. The phase shift in the
chiral limit clearly extends down to $s=0$, and since this is the
S-wave, the strength turns on relatively quickly. The physical case
needs to vanish at the physical threshold. These requirements
naturally yields a larger phase shift for the chiral case throughout
much of the physical region. At high energies, we see a much smaller
difference between the chiral limit and the physical case. This is
what should be expected, as the pion mass should make less of a
difference at the higher energies.The Omnes function follows
directly from the phase shifts and also has recognizable features.
This is shown in Fig \ref{chiral Omnes1}.

These results are encouraging for the reliability of the method. The
place where we have the least theoretical control is the upper
energy end of the energy region. The IAM does not produce a much
variation at these energies, so that most of the variation that we
find then comes from physics at lower energies. Moreover, the phase
shifts only enter our calculation through the Omnes function, and
the Omnes function is small at high energy. The true Omnes function
from the CGL phase shifts is yet smaller, so that the IAM does
slightly overemphasize the higher energies. However, since there is
not much mass variation at these energies, this is not a serious
flaw. Overall, despite the approximate nature of the IAM method,
there is no reason to doubt the general features of the resulting
phase shift and hence the Omnes function.

\section{Dispersive chiral analysis}

 In this section, the evaluation of the contact interaction
will be presented. The key features of the scalar channel (i.e. the
``sigma'') emerge, as reported in \cite{sigma}. Moreover, the chiral
limit will induce a shift in the scalar coupling that again emerges
mostly from low energy two-pion exchange. In performing this
comparison, I have not varied the strange quark mass, keeping it
fixed at its physical value.

In addition to the explicit dependence on the pion mass in the
chiral amplitudes and the Omnes function, there is also implicit
dependence on the pion mass contained in the parameters $g_A$ and
$F_\pi$.  The dependence has the form\cite{gasser, implicit}
\begin{eqnarray}
F_\pi &=& F_0 \left[ 1- \frac{1}{16\pi^2 F_0^2} m_\pi^2 \left( \ln \frac{m_\pi^2}{m_{ph}^2} + \bar{l}_4 \right)\right] \nonumber \\
g_A &=&  g_0 \left[ 1 - \frac{2g_0^2 +1}{16\pi^2 F_0^2} m_\pi^2 \ln
\frac{m_\pi^2}{m_{ph}^2} - \frac{g_0^2 m_\pi^2}{16\pi^2 F_0^2} +
\frac{4 m_\pi^2 \bar{d}_{16} }{g_0}\right]
\end{eqnarray}
with $\bar{l}_4 = 4.4 \pm 0.2$ and $\bar{d}_{16} = -1.0 \pm 0.7$.
While the mass dependence of the pion decay constant is fairly well
constrained, the corresponding result in $g_A$ is less well
understood. This will be a significant component of our final error
estimate.

The chiral parameters $c_{1,2,3}$ are determined from a chiral
analysis of pion-nucleon scattering\cite{bernard, lattice}. When
used without matching to the Omnes function at second order in
$m_\pi^2$, as is the case both in the $\pi N$ analyses and in this
paper, there is an ambiguity as to whether the fitted parameter
should be identified with $c_3$ itself or $c_3 \Omega(2m_\pi)$ which
is the threshold value of the coupling. Since $\Omega(2m_\pi)\sim
1.25$, this makes a difference in the result. The spectral function
is dominantly determined by the parameter $c_3$, which is however
not very well determined, with $c_3=- 4.7^{+1.2}_{-1.0} $~GeV$^{-2}$
being the quoted value. If we used instead the constraint
$c_3\Omega(2m_\pi)=- 4.7^{+1.2}_{-1.0} $~GeV$^{-2}$, we would choose
$c_3= -3.7^{+1.0}_{-0.8}$. The overall magnitude of $G_S$ does
depend sensitively on the choice of $c_3$, see \cite{sigma}. However
the ratio of $G_S$ in the chiral limit to that of the physical limit
is almost completely insensitive to $c_3$, as will be commented on
below. Therefore, I will only quote the ratio as my final result.
Although I have explored a wide range of $c_3$ values, the figures
were produced using $c_3=- 4.0 $~GeV$^{-2}$, a result intermediate
between the two choices mentioned above and consistent with both
within errors. The other parameter choices used were $c_1=-0.64
$~GeV$^{-2}$ (i.e. $c_1\Omega(2m_\pi) = -0.8$), and $c_2=3.3
$~GeV$^{-2}$. I have explored the sensitivity of the results to
these parameters, and both the magnitude and the ratio are
insensitive to reasonable variations.

At this stage we can put the pieces together. The chiral amplitudes
of Eq. \ref{spectral} contain both explicit and implicit mass
dependence, displayed above. The Omnes function has further
dependence. The resulting integrand in the spectral sum rule for
$G_S$ has been previously shown in Fig. 1, both for the physical
case and for the chiral limit.

As described in \cite{sigma}, the shape of the spectral function in
the physical case provides a reasonable representation of the sigma
that traditionally appears in the nuclear potential. Moreover, for
reasonable values of the chiral parameters, in particular $c_3$
which has the greatest impact, the physical value of $G_S$ can be
reproduced. In this way we can understand the nature of ``sigma
exchange'' in the nuclear force. The sigma is not easily understood
in terms of the quark and gluon degrees of freedom of QCD. It is
also not a conventional resonance, i.e. a pole in the scattering
amplitude close to the real axis with the scattering phase passing
through 90$^o$ near the pole. In $\pi\pi$ scattering, the phase
shift is not large near the sigma and careful analysis\cite{ccl}
reveals a pole on the second sheet which is very far from the real
axis. Other approaches have also described an effective sigma
exchange without invoking a specific resonance\cite{meissnersigma,
grein, oset, goldman}. In the present case we have a very simple set
of ingredients, each if which is required within its respective
domain. Even though the amplitude is not conventionally resonant at
500-600~MeV, the broad peaking of the spectral function reproduces
an effect very similar to a sigma resonance.

Since we have analytic control over the ingredients of this
calculation we may use this formalism to take the pion mass to zero.
The result in the chiral limit has a similar high energy behavior
and the magnitude there is only slightly increased. What is most
striking is the enhancement at low energy. Much of this is purely
kinematic. The threshold in the chiral limit extends down to zero
energy and the chiral predictions for the spectral function develops
quickly.

To calculate the ratio of the scalar strength in the chiral limit to
the physical limit, one compares the integral under the two curves
of Fig. 1. The extra weight under the spectral integrand causes the
value of $G_S$ to increase significantly. Overall I find
\begin{equation}
\eta_{S,~ch} \equiv\frac{G_S|_{chiral}}{G_S|_{physical}} = 1.37
\end{equation}
This is our primary result. It will be applied to the nuclear
interaction in a subsequent section.

Let me attempt to assess the uncertainties in this result. It is
clear that this calculation is not straight chiral perturbation
theory, and hence does not share the rigor of that method. I have
modeled the moderate energy behavior in the dispersive sum rule, and
neglected all truly high energy effects beyond 1 GeV. We clearly
have no control over the effects at very high energy. However, there
is no indication either phenomenologically or theoretically that
they are very important in the determination of $G_S$, nor that they
would have significant dependence on the pion mass. We can look at
the low and moderate energy uncertainties in more detail.

Most of the the shift in the scalar coupling comes from the low
energy region. The threshold behavior is an unambiguous feature -
the chiral limit amplitudes must extend down to zero energy and the
chiral expansion becomes exact there. Even at moderate energies, the
IAM representation of the pion phase shifts should be quite good.
The greatest uncertainty in this region is the lack of understanding
of the chiral behavior in the pion coupling to nucleons, as the
leading spectral representation at low energy has a factor of
$g_A^2$ . This effect is quantifiable. The uncertainty in the chiral
parameters $d_{16}$ and $\bar{l}_4$ corresponds to an uncertainty of
$\pm 0.07$ in our final result. Perhaps this uncertainty may be
reduced in the future by lattice calculations. There could be
additional $m_\pi^2/(1 {\rm GeV})^2 \sim 0.02 $ corrections to the
$c_{1,2,3} $ vertices in addition to that modeled by the Omnes
factor, so there are clearly other uncertainties at the order of
several percent.

The overall strength of $G_S$ depends most heavily on the chiral
parameter $c_3$. However, the ratio of the chiral limit to the
physical case has only a remarkably tiny variation with this
parameter. Throughout the whole range of $c_3$ considered above, we
find the ratio changes by less than 1\%. Thus the uncertainty due to
the magnitude of this parameter is insignificant compared to other
uncertainties.

\begin{figure}[ht]
 \begin{center}
   \begin{minipage}[t]{.07\textwidth}
    \vspace{0pt}
    \centering
    \vspace*{0 in}
    \hspace*{200pt} {$r$~(GeV$^{-1}$)}\\
    \vspace*{60 pt}
    \hspace*{65pt}{$V(r)$} \\
  \end{minipage}%
  \begin{minipage}[t]{0.93\textwidth}
    \vspace{05 pt}
    \centering
\hspace{-0.0 in}\rotatebox{-0}
{\includegraphics[width=0.6\textwidth,height=!]{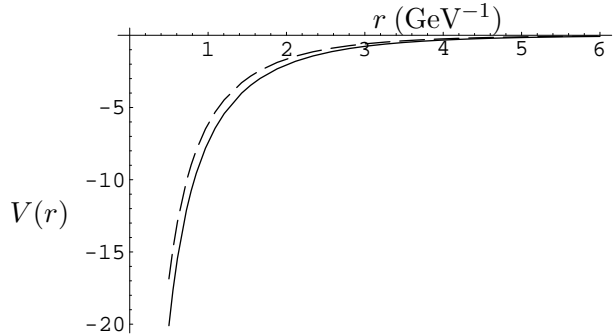}}\\
    ${} \vspace*{-0pt}$
  \end{minipage}
 \end{center}
 \caption{The attractive scalar-isoscalar potential in the physical
 case (dashed line) and in the chiral limit (solid line).}
 \label{chiral Vr}
\end{figure}

One might worry that the treatment of two-pion exchange as a
potential, or the use of a contact interaction instead of a
potential, would break down in the chiral limit because of the
massless pions. This could be the case if the very long range tail
of the potential was significantly modified. Within this
calculation, there is no evidence of a problem. The spectral
function also can be used to predict a spatial potential $V(r)$. The
physical result and the chiral limit result are shown in Fig.
\ref{chiral Vr}. The shape of the chiral result is quite similar to
the physical limit.

The moderate energy effects beyond the realm of straight chiral
perturbation theory are less quantifiable, but we have reason to
believe that they are not large. In contrast to the low energy
effects, there is no reason to expect that these effects are
enhanced over the usual expectation of being of order $m_\pi^2
/\Lambda^2$. Indeed we find a reasonably small shift in the spectral
function at high energy, so that our explicit calculation is
consistent with this. However, I find it hard to defend that high
energy shift as a solid prediction of the method. If I were to
neglect all shifts above 500 MeV, the main result for $\eta_S$ would
be about 5\% smaller.

We can perform a further test by adding some expected higher energy
effects. One uncertainly that we are able to explore numerically
involves the energy dependence of the pion vertices. The two pion
vertex would presumably have some formfactor associated with the
leading coupling, $c_3$, which would generate a dependence of the
vertex on the momentum transfer. I have studied this by creating an
energy dependent vertex through the substitution
\begin{equation}
c_3 \to \frac{c_3}{(\mu^2+m^2)^n}
\end{equation}
As mentioned in \cite{sigma}, this modifies the shape of the
spectral function somewhat, and changes the integral that determines
$G_S$. However, the change is modest enough that the desired value
can still be obtained by changing the fit value of $c_3$ within the
allowed range. In this case there is a modest increase
chiral/physical ratio of $G_S$. For example with $n=1$ and
$m=1200$~MeV there is a 29\% decrease in the predicted value of
$G_S$ at fixed $c_3$. However, for the ratio $\eta_{S,~ch}$ this
amounts to a 7\% increase leading to the result $\eta_{S,~ch}=1.46$.
The decrease in $G_S$ and increase in $\eta_S$ is readily
understood. The formfactor will decrease the contribution at higher
values of $\mu$, but leave unchanged the threshold effects at small
$\mu$. The former feature will decrease the physical value of $G_S$,
but the new ingredients enhancing the chiral limit will be
unchanged.

Additionally, in the section describing the chiral slope below, I
will mention a puzzle concerning the lack of understanding of the
nonanalytic behavior. While it is numerically small, this is also a
feature which must somehow be generated from the higher energy
portion of the spectral function.

I would summarize this discussion of the uncertainties by giving a
final result of
\begin{equation}
\eta_{S,~ch} \equiv\frac{G_S|_{chiral}}{G_S|_{physical}} = 1.37 \pm
0.10
\end{equation}

\section{The vector channel}

Also contributing to the central force is the dispersive channel
involving three pions in an $I=0$ state -i.e. the $\omega$ exchange
channel. In this section, I will estimate the modification of this
strength in the chiral limit. The change is very slight, and the net
effect on nuclear binding is so small that it is well below even the
uncertainty of the result in the scalar channel. The reason for this
is clear - the $3\pi$ channel is very small at threshold and hence
the threshold sensitivity that the scalar channel exhibited is
absent. The chiral modification has the natural size of order
$m_\pi^2 / ({\rm 1 GeV})^2$.

In the dispersion relation
\begin{equation}
G_{V} =\frac{2}{\pi}\int_{3m_\pi}^{\infty} \frac{d\mu}{\mu}
~{\rho}_{V}(\mu^2)
\end{equation}
the spectral function ${\rho}_{V}(\mu^2)$ is small at low energy. In
chiral perturbation theory it arises first as the cut in a two loop
diagram, and hence is higher order in the energy expansion. There is
also a kinematic suppression since the 3 pions are in a total
spin-one state. Both theoretically and phenomenologically there
seems to be little strength in this channel, see e.g.
\cite{bernard}, until one reaches the energy of the $\omega(783)$,
which is a narrow resonance coupled essentially entirely to three
pions.  After the $\omega(783)$, there is not another three pion
resonance until well over 1~GeV. It is then a reasonable
approximation to take the vector spectral function dominated by the
narrow resonance $\omega(783)$.

We need to understand how the properties of the $\omega(783)$ - i.e.
its mass and coupling - are modified in the chiral limit. We can
estimate its mass in the chiral limit from the SU(3) breaking in the
other vector mesons. Replacing one of the light quarks by a strange
quark yields the $K^*(890)$ and replacing two of them produces the
$\phi(1040)$. The vector mesons empirically obey an equal spacing
rule. This leads to a simple linear interpolation formula as a
function of the light quark mass $\hat{m}$:
\begin{equation}
m_V = m_\omega + (m_\phi - m_\omega)
\frac{\hat{m}-\hat{m}_{phys}}{m_s-\hat{m}_{phys}}
\end{equation}
The chiral limit is found with $\hat{m}=0$, resulting in
\begin{equation}
m_{chiral}= 783 -  (1040 - 783)
\frac{\hat{m}_{phys}}{m_s-\hat{m}_{phys}} = 770
\end{equation}
This is a 2\% shift in the mass.

One can address the quark mass dependence of the vector coupling to
nucleons again through the SU(3) breaking pattern plus data. The
SU(3) breaking of the NNV couplings themselves is not known well
enough experimentally to be of direct use. However an indirect
method can be employed. There is a phenomenologically successful
model for these coupling, called vector meson dominance, that
related the nucleon's NNV coupling $g_v$ to the coupling of the
vector meson to a photon $f_v$, defined via
\begin{equation}
<V | J_\mu|0> = c_v f_v m_v^2 \epsilon_\mu
\end{equation}
where $J_\mu$ is the electromagnetic current and $c_v $ are known
Clebsch-Gordon coefficients. The VMD relation is simply $g_v =
f_v^{-1}$, and can be motivated/derived from a dispersion relation
argument for the electromagnetic current of the nucleon.
Phenomenologically VMD works very well. Since there is experimental
evidence on the SU(3) breaking of $f_v$, we can use that plus VMD as
an estimate of the quark mass dependence of $g_v$. I have performed
a fit to the data involving the neutral vector mesons
$\rho,~\omega,~\phi$ again assuming a linear interpolation. The
result is a dependence of the form
\begin{equation}
g_v = g_{v0} [ 1 + b_{gv} (m_\pi^2 - m_{ph}^2)]
\end{equation}
with $b_{gv} = 0.57 \pm 0.22$ GeV$^{-2}$. The error bar comes from
using different data in the evaluation of the SU(3)breaking. The
central estimate is then
\begin{equation}
\frac{g_v|_{chiral}}{g_v|_{physical}} =0.99
\end{equation}

We can combine the vector mass and coupling to determine the shift
in the parameter vector coupling $G_V = g_V^2/m_V^2$. This yields
\begin{equation}
\eta_{V,~ch} \equiv \frac{G_V|_{chiral}}{G_V|_{physical}} = 0.995
\pm 0.020
\end{equation}
The variation of both the mass and the coupling constant were mild
compared to the dependence that we will estimated for the scalar
channel. Moreover, both the mass in the denominator and the coupling
constant in the numerator decreased, leading to a very small net
effect in the ratio. The error bar quoted is a generous estimate. In
any case, the uncertainty in this coupling is significantly smaller
that that in the scalar coupling.

\section{Estimate of the changes in nuclear binding energy}

Our results clearly have the strength of the attractive scalar
coupling being significantly increased in the chiral limit. As I
argued in Sec. II, in effective field theory the most consistent
procedure would be to treat the long range components of the two
pion exchange dynamically, most likely with a form of a cutoff, and
then add the contact interactions to account for the rest of the
interaction. This has not yet been done for heavy nuclei. However,
the use of the contact interactions in point-coupling calculations
of binding provide an estimate of the shift in the binding that is
appropriate for the present state of the art.

Not all possible contact interactions play a significant role in
nuclear binding. The dominant ingredients in the binding of heavy
nuclei have been elucidated by Furnstahl, Serot and
co-workers\cite{Furnstahl:1999rm, otherbinding}. The results can be
extracted from Fig. 1 and Fig. 2 of \cite{Furnstahl:1999rm}. It is
clear that the dominant effects are the scalar and vector
contributions, to the leading power of the density. Other
interactions play reduced roles, although for a complete
understanding of the binding about a half-dozen contact interactions
are required. I will consider only the dominant isoscalar-scalar and
isoscalar-vector interactions, and in practice it is only the scalar
coupling that has a significant dependence on the pion mass.

Using Ref.~\cite{Furnstahl:1999rm}, one can read off the effects of
the different contact terms\footnote{I thank Dick Furnstahl and
Brian Serot for assistance in understanding these numbers.} I
parameterize the results in terms of the strengths of the contact
interactions, normalized to their physical values, defining
 \begin{eqnarray}
\eta_S &=& \frac{G_S}{G_S|_{physical}}
\nonumber \\
\eta_V &=& \frac{G_V}{G_V|_{physical}}
\end{eqnarray}
The contributions to the binding energy numbers for $^{16}O$ (in
MeV) are
\begin{equation}
{{\rm B.E.}\over A} \sim -82 \eta_S + 44\eta_V + 30
\end{equation}
The first two terms are the effects of the scalar and vector
isoscalar interactions. The third term is the sum of 4 other smaller
contributions to the binding energy and kinetic energy
contributions. There is in addition the coulomb energy and a small
center of mass correction. For $^{208}Pb$, the result is
\begin{equation}
{{\rm B.E.}\over A} \sim -104 \eta_S + 57\eta_V + 36
\end{equation}

The results of these calculations can be generalized to other nuclei
by a parameterization that resembles the semi-empirical mass
formula. For local interactions, because the nuclear density is
nearly constant in the central region one expects that the binding
energy will have a dependence volume, which in turn is proportional
to the number of particles $r^3\sim A$, and that interactions that
occur near the nuclear surface would have a modified result
proportional to the number of nucleons near the surface,$r^2\sim
A^{2/3}$. This suggests that binding effects can be parameterized in
terms of behavior in $A$ and in $A^{2/3}$. Using the results for
nuclear matter and for specific nuclei, we find a good fit of the
form
\begin{equation}
{{\rm B.E.}\over A} = -(120 -\frac{97}{A^{1/3}}) \eta_S + (67 -
\frac{57}{A^{1/3}} )\eta_V  +{\rm residual ~terms}
\end{equation}

Our results from the previous sections can be summarized as
\begin{eqnarray}
\eta_{S,~ch} &=& \frac{G_S|_{chiral}}{G_S|_{physical}} = 1.37 \pm
0.10
\nonumber \\
\eta_{V,~ch} &=& \frac{G_V|_{chiral}}{G_V|_{physical}} = 0.995
\end{eqnarray}
These numbers produce
\begin{equation}
{{\rm B.E.}\over A}|_{chiral} = -38 \pm 8~~{\rm MeV}
\end{equation}
for $^{16}O$, compared to the physical result of $-8 $~MeV. For
$^{208}Pb$ the corresponding results are  $-49 \pm 10 $ MeV in the
chiral limit compared to $-11$~MeV in the physical case. Finally for
the general paramaterization, the results are suggest a shift in the
binding energy
\begin{equation}
\Delta {{\rm B.E.}\over A}|_{chiral} = -(44 - \frac{36}{A^{1/3}})
\end{equation}

These shifts are larger than we would naively expect. There are two
ingredients in generating this magnitude. First is the large shift
in the scalar contact interaction $G_S$. I have explained at length
above why this large shift occurs and why it should be considered
natural. The other ingredient is that there is a strong cancelation
between the scalar and vector terms in the expression for the
binding energy. The usual 10 MeV/nucleon binding energy is the
difference of two significantly larger numbers. This is
understandable from the meson exchange potential description. Scalar
exchange provides a strong attractive potential. Vector meson
exchange provides the repulsive short range interaction and tends to
oppose binding. The fact that the 10 MeV of binding energy is far
below all other hadronic scales of QCD comes from the competition of
these two opposing effects. When we go to the chiral limit and make
the attractive interaction significantly stronger while leaving the
repulsive one unchanged, the near cancelation between these
competing effects becomes less pronounced and the percentage shift
is large.

\section{The chiral slope parameter}

Another way to present the results of this calculation is as the
slope in the scalar coupling as one deviates from the chiral limit.
The leading terms in an expansion around the chiral limit are given
by
\begin{equation}
G_S = G_{S0}(1 + d_S m_\pi^2) + F_S m_\pi^3 +...
\end{equation}
where $G_{S0}$ is the result in the chiral limit, and the
non-analytic term is \cite{brockmann}
\begin{equation}
F_S = -\frac{3g_A^2}{16\pi F_\pi^4}(6c_1-5c_3)
\end{equation}
The effect of the $F_S$ term is relatively small, i.e. $F_S
m_\pi^3/G_{S0} \sim 0.09$ .

It should be pointed out that the dispersion relation does not
exactly reproduce the correct $m_\pi^3$ dependence of the chiral
coupling. The cubic mass dependence arises in the dispersive
calculation from the threshold behavior of the chiral amplitude. Use
of the chiral representation for the spectral function given in Eq.
\ref{rho1} leads to a cubic term that is
\begin{equation}
F^{(disp)}_S = -\frac{3g_A^2}{16\pi F_\pi^4}(6c_1-\frac{11}{3}c_3)
\end{equation}
I have checked both the spectral function and the real part of the
loop calculation. This has nothing to do with the unitarization
procedure and cannot be changed by higher order corrections to the
threshold behavior. While the effect is numerically small, it
represents a puzzle for which I do not understand the
resolution\footnote{I thank E. Epelbaum, U. Mei{\ss}ner, N. Kaiser
and Barry Holstein for discussions about this issue}. It is
fortunate that this nonanalytic term is numerically small and the
effect of the disagreement has only a 2\% effect on the
determination of the slope.

I have extracted the chiral slope by the evaluation of the
dispersion integral at several values of the pion mass and
performing a fit using terms in $m_\pi^2$ and  $m_\pi^4$ as well as
the known nonanalytic term in $m_\pi^3$. The result of the fit
yields the slope
\begin{equation}
d_S = -(17\pm 5)~{\rm GeV}^{-2} = -\frac{0.31 \pm 0.08}{m_\pi^2}
\end{equation}
The negative sign is indicative of the the fact that the scalar
coupling is smaller in the physical case than it is in the chiral
limit. This slope is relatively large for the reasons discussed
above. Again, because the ratios of scalar couplings are predicted
better than the absolute values, the parameter $d_S$ is much better
determined in this calculation than $D_S\equiv d_S G_{S0}$. As seen
in the previous section, the slope of the combination of the two
isoscalar effects in the overall nuclear central force is yet larger
because of the partial cancelation of the scalar and vector
channels.

\section{Summary and discussion}

I have used the chiral results for the low energy behavior of the
dispersion relation for the scalar contact interaction, supplemented
with a representation for the Omnes function, in order to calculate
the strength of the central nuclear force in the chiral limit. In
the results, the largest modifications are seen to come from the low
energy amplitudes, which extend down to zero energy in the chiral
case. While there are some uncertainties in the final result, most
notably from the uncertainty in the chiral extrapolation of $g_A$,
the key ingredients in the calculation appear clear. We also
understand qualitatively why the effect is relatively large.

Since this has been an attempt to calculate the full scalar
coupling, I have had to model the high energy contributions. While I
have argued that this modeling has not introduced large effects in
the final results, there is also a way to use this method with the
full rigor of chiral perturbation theory. To do this, one would use
the present calculation dynamically and in addition introduce a
residual contact interaction to account for high energy effects
which have been misrepresented in the present calculation. This
extra parameter would presumably be small and would not be expected
to have much mass variation. This variant of the Wilsonian scheme
described in Sec. 2 would then be a way to implement the effective
field theory treatment while capturing the main dynamical results of
the present calculation.

There exists in the literature several other works which discuss the
nuclear interaction in the chiral limit\cite{meissnerchiral,
savagechiral, lattice} within the context of chiral perturbation
theory. The present work is different because it is a dynamical
attempt to calculate the mass variation in the scalar channel.
Previous work has been both more general because this variation was
parameterized a chiral coefficient, and also more limited because
this coefficient could only be guessed at. In practice the mass
variation found in this paper is larger than the estimates of
\cite{meissnerchiral, savagechiral}, and has a well-determined sign.
Another difference is that the previous works focussed on
few-nucleon systems. This requires the addition of single pion
exchange, which obviously also has a significant change in the
chiral limit. My contribution here is only on the scalar sector -
hence I have limited my comments on binding to the heavy nuclear
case for which one pion exchange is not important. However, it is
clear that the results of this paper would push the binding of
few-nucleon systems in the direction of {\em greater} binding. It
would be interesting to revisit the few-nucleon case with either the
parameter $d_S$ calculated in the present work or with a full
Wilsonian treatment dynamically including the effects of low energy
two-pion exchange.

\section*{Acknowledgement} I would like to thank especially Thibault
Damour for many conversations during the course of this work. In
addition I have benefited from discussions with U. von Kolck  R.
Furnstahl, B. Serot, E. Epplebaum, U. Mei{\ss}ner, N. Kaiser and
Barry Holstein. I also appreciate the hospitably the IHES where most
of this work was performed. This work has been partially supported
by the U.S National Science Foundation.

\end{document}